\documentclass[twocolumn]{aastex6}
\pdfoutput=1 
\usepackage[T1]{fontenc}
\usepackage{amsmath,amstext}
\usepackage{apjfonts} 
\usepackage[figure,figure*]{hypcap}
\usepackage{microtype}
\usepackage{caption}
\usepackage{tablefootnote}
\usepackage{etoolbox}
\newcommand{\repeater}{FRB121102} 

\begin{document}
\shortauthors{Eftekhari et al.}
\shorttitle{A Radio Source Coincident with the Superluminous Supernova PTF10hgi}

\title{A Radio Source Coincident with the Superluminous Supernova PTF10\MakeLowercase{hgi}:  Evidence for a Central Engine and an Analogue of the Repeating \repeater{}?}
\author{{T.~Eftekhari}\altaffilmark{1}}
\author{E.~Berger\altaffilmark{1}}
\author{{B.~Margalit}\altaffilmark{2}$^*$}\altaffiliation[*]{NASA Einstein Fellow}
\author{P.~K.~Blanchard\altaffilmark{1}} 
\author{L.~Patton\altaffilmark{1}}
\author{P.~Demorest\altaffilmark{3}} 
\author{P.~K.~G.~Williams\altaffilmark{1}}
\author{S.~Chatterjee\altaffilmark{4}} 
\author{J.~M.~Cordes\altaffilmark{4}}
\author{R.~Lunnan\altaffilmark{5}}
\author{B.~D.~Metzger\altaffilmark{6}}
\author{M.~Nicholl\altaffilmark{7,8}}

\altaffiltext{1}{Center for Astrophysics | Harvard \& Smithsonian, Cambridge, MA 02138, USA}
\altaffiltext{2}{Astronomy Department and Theoretical Astrophysics Center, University of California, Berkeley, Berkeley, CA 94720, USA}
\altaffiltext{3}{National Radio Astronomy Observatory, Socorro, NM 87801, USA}
\altaffiltext{4}{Cornell Center for Astrophysics and Planetary Science and Department of Astronomy, Cornell
University, Ithaca, NY 14853, USA}
\altaffiltext{5}{The Oskar Klein Centre \& Department of Astronomy, Stockholm University, AlbaNova, SE-106 91 Stockholm, Sweden}
\altaffiltext{6}{Department of Physics and Columbia Astrophysics Laboratory, Columbia University, New York, NY 10027, USA}
\altaffiltext{7}{Institute for Astronomy, University of Edinburgh, Royal Observatory, Blackford Hill, Edinburgh EH9 3HJ, UK}
\altaffiltext{8}{Birmingham Institute for Gravitational Wave Astronomy and School of Physics and Astronomy, University of Birmingham, Birmingham B15 2TT, UK}

\begin{abstract}
We present the detection of an unresolved radio source coincident with the position of the Type I superluminous supernova (SLSN) PTF10hgi ($z=0.098$) about 7.5 years post-explosion, with a flux density of $F_\nu(6\,\,{\rm GHz)}\approx 47.3\ \mu Jy$ and a luminosity of $L_\nu(6\,\,{\rm GHz})\approx 1.1\times 10^{28}$ erg s$^{-1}$ Hz$^{-1}$.  This represents the first detection of radio emission coincident with a SLSN on any timescale. We investigate various scenarios for the origin of the radio emission: star formation activity, an active galactic nucleus, and a non-relativistic supernova blastwave. While any of these would be quite novel if confirmed, none appear likely when taken in context of the other properties of the host galaxy, previous radio observations of SLSNe, and the general population of hydrogen-poor SNe. Instead, the radio emission is reminiscent of the quiescent radio source associated with the repeating \repeater{}, which has been argued to be powered by a magnetar born in a SLSN or LGRB explosion several decades ago. We show that the properties of the radio source are consistent with a magnetar wind nebula or an off-axis jet, indicating the presence of a central engine.  Our directed search for FRBs from the location of PTF10hgi using 40 min of VLA phased-array data reveals no detections to a limit of $22$ mJy ($10\sigma$; 10 ms duration). We outline several follow-up observations that can conclusively establish the origin of the radio emission.
\end{abstract}

\keywords{radio continuum: transients}

\section{Introduction}
\label{sec:intro}

Fast radio bursts (FRBs) are bright, GHz frequency, millisecond duration pulses with dispersion measures (DMs) well in excess of Galactic values, pointing to an extragalactic origin \citep{Lorimer2007}. The discovery of the repeating \repeater{} \citep{Spitler2014,Spitler2016} enabled the first precise localization of an FRB \citep{Chatterjee2017}, which in turn led to the identification of the host as a star forming low metallicity dwarf galaxy at $z=0.193$ \citep{Tendulkar2017}.  The nature of the host, coupled with the discovery of a parsec-scale ($\lesssim 0.7$ pc), persistent radio source coincident with the bursts \citep[$\lesssim$40~pc;][]{Marcote2017}, have prompted theories suggesting that FRBs are powered by decades-old millisecond magnetars born in superluminous supernova (SLSN) and/or long gamma-ray burst (LGRB) explosions \citep{Murase2016,Piro2016,Metzger2017,Nicholl2017c}. Within this framework, we expect the locations of at least some known SLSNe and/or LGRBs to produce FRBs and to be accompanied by quiescent radio sources on roughly a decade timescale post-explosion, as the expanding ejecta become transparent to free-free absorption at GHz frequencies \citep{Omand2017,Margalit2018a}.

To test this prediction, we recently carried out VLA and ALMA searches for quiescent radio/mm sources in a volume-limited sample of SLSNe and LGRBs (Eftekhari et al.~in prep.). In the VLA observations we simultaneously searched for FRBs from the same locations using phased-array observations. We note that the same data can also probe other interesting aspects of SLSNe and their host galaxies, namely the presence of obscured star formation, an active galactic nucleus (AGN), interaction of the SN blastwave with circumstellar material, and an off-axis jet. The latter possibility, in addition to the scenario of an \repeater-like quiescent source, would provide direct evidence for a central engine in SLSNe; such direct evidence is currently lacking (e.g., \citealt{Coppejans2018,Bhirombhakdi2018}) despite the fact that modeling of SLSN light curves, and observations of their nebular spectra, point to a magnetar central engine (e.g., \citealt{Nicholl2017b,Nicholl2018}).

\begin{figure*}
\begin{center}
\includegraphics[width=0.95\textwidth]{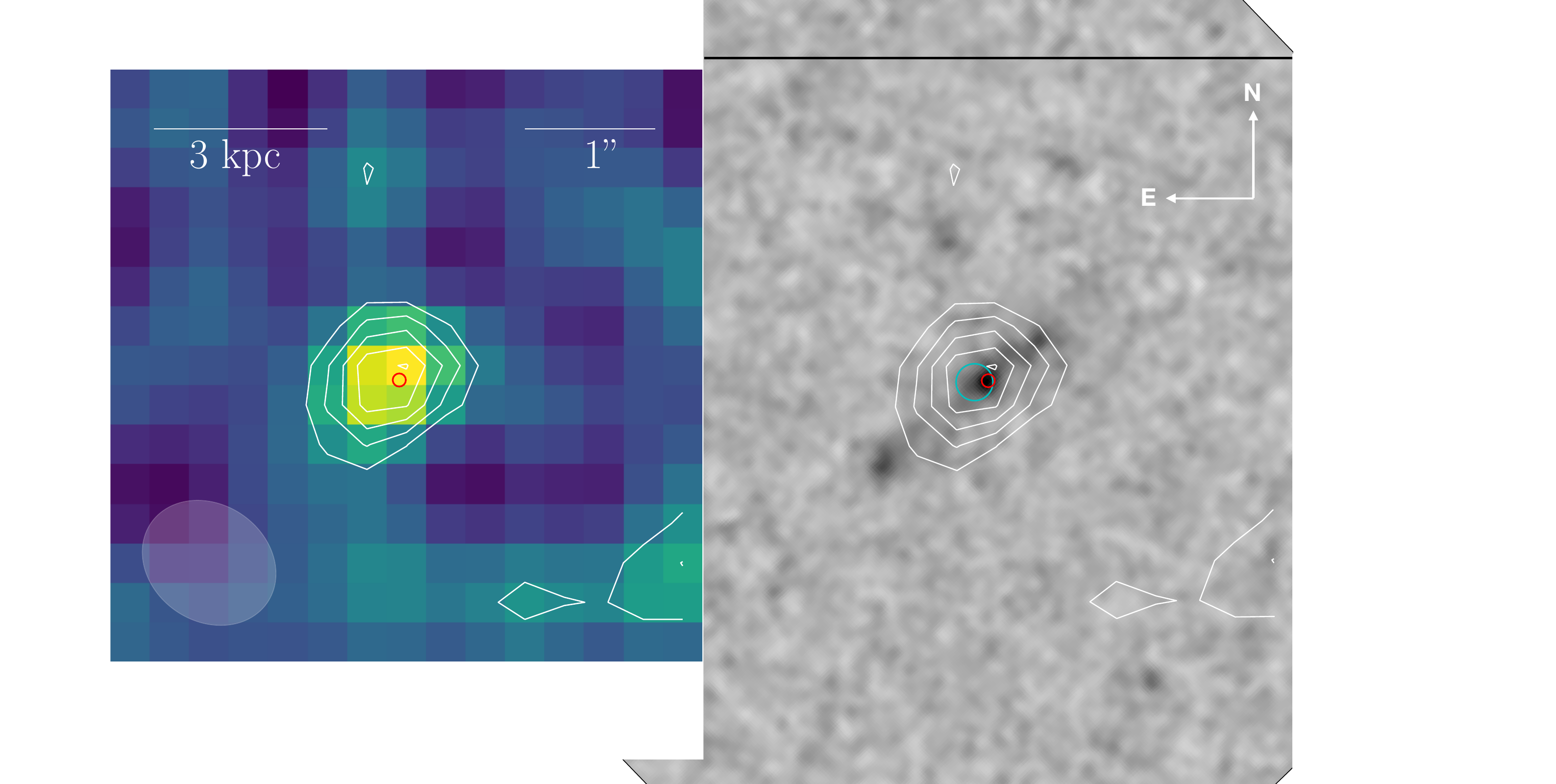}
\end{center}
\caption{\textit{Left:} Radio continuum map from VLA 6 GHz (C band) observations of PTF10hgi. Contours correspond to $-2$, 2, 3, 4, 5, and 6 times the root-mean-square noise of the image. The synthesized beam ($1.1'' \times 0.9''$) is shown in the lower left corner. Also shown is the optical position of PTF10hgi (red circle; $2\sigma$). \textit{Right}: Near-UV image of the host galaxy of PTF10hgi from \textit{HST}/WFC3 with radio contours and the fitted position of the radio source (cyan circle; $2\sigma$) overlaid. Details of the astrometry are provided in \S\ref{sec:astrom}.}
\label{fig:image}
\end{figure*}

Here we report the VLA detection of an unresolved radio source coincident with the location of the SLSN PTF10hgi ($z=0.098$; \citealt{Inserra2013,Perley2016a,DeCia2018}) about 7.5 years post-explosion. PTF10hgi was classified as a Type I SLSN by \citet{Quimby2010} because its maximum light spectra were dominated by a blue continuum, with no obvious emission or absorption lines, similar to many events in this class. However, \citet{Quimby2018} recently found that by $\sim 1$ month after peak, the cooler spectrum showed broad H and He lines in addition to the usual Fe, Ca, O and Mg lines at this phase. Although PTF10hgi is unique in this respect, its other properties, including colors, peak luminosity and light curve shape, are consistent with other Type I SLSN.

This represents the first detection of radio emission coincident with a known SLSN on any timescale (e.g., \citealt{Coppejans2018,Hatsukade2018}). We investigate the various possible origins of the radio emission --- star formation activity, AGN, and SN blastwave --- and show that none are likely, although in each scenario such an origin would represent an exciting and novel result.  Instead, if supported by additional observations, the radio source may represent the first detection of non-thermal emission from a SLSN engine -- either in the form of an off-axis jet or a magnetar wind nebula -- providing compelling evidence for the millisecond magnetar model of SLSNe, as well as for a connection between repeating FRBs (and perhaps all FRBs) and millisecond magnetars born in SLSN explosions. We present the observations in \S\ref{sec:obs}, present and discuss various models for the radio emission in \S\ref{sec:origin}, and summarize with a discussion of future observations in \S\ref{sec:conc}.

\section{Observations and Data Analysis}
\label{sec:obs}

\subsection{VLA Continuum Observations}
\label{sec:vla}

We observed the location of PTF10hgi with the Karl G.~Jansky Very Large Array (VLA), in the B configuration, on 2017 December 15 UT. We used the C band wideband continuum mode with the 8-bit samplers configured to two basebands with center frequencies of 5 and 7 GHz and 1 GHz bandwidth each. The total on-source time of the observations was 40.5 min.  We applied standard calibration techniques using 3C286 for bandpass and flux density calibration and J1658+0741 for complex gain calibration.  We note that $8$ antennas were offline for our observations.

We processed the data in the Common Astronomy Software Application (CASA) software package \citep{McMullin2007} using standard imaging techniques. We use the CASA task \texttt{CLEAN} to Fourier invert the complex visibilities and deconvolve the dirty image. The image is gridded to a size of 3000 pixels at a scale of 0.3 arcsec per pixel using multi-frequency synthesis (MFS; \citealt{Sault1994}) and $w$-projection with 128 planes \citep{Cornwell2008}. We fit for the flux density and source position using the \texttt{imtool} program as part of the \texttt{pwkit}\footnote{Available at https://github.com/pkgw/pwkit.} package \citep{Williams2017}. We note that we do not present an analysis of polarization given that we do not have proper polarization calibration. Furthermore, at our sensitivity of $\sim$7 $\mu$Jy, a confident detection of polarization would not be possible, unless it is at the 100\% level.

We identify a point source with a flux density of $F_\nu = 47.3\pm 7.1 \mu$Jy ($6.7\sigma$) at R.A.=${\rm 16^h37^m47^s.071}$, decl.=$+06^\circ12'31''.88$ (J2000) with an uncertainty of $0.14''$ in each coordinate. The uncertainty on the flux density includes the uncertainty on the source size and position. We also fit a Gaussian to the observed emission in the image plane using the CASA task \texttt{imfit} and find that the emission is consistent with a point source and hence unresolved. We further image the two sidebands separately to constrain the spectral index of the source and find $\alpha=0.85\pm 1.65$ ($F_\nu\propto \nu^\alpha$). An image of the field, centered on the location of the radio source, is shown in Figure~\ref{fig:image}. 

\subsection{VLA Phased-Array Observations} 
\label{sec:phased}

In addition to the standard continuum observations, we also obtained simultaneous phased-array observations to search for individual ms-duration bursts from PTF10hgi. The summed phased-array data were recorded with 2 GHz total bandwidth with 256 $\mu$s time resolution and 2 MHz channels. The raw filterbank files are divided into two channelized time series of 1 GHz bandwidth each with center frequencies of 5 and 7 GHz. We searched each file for RFI using \texttt{PRESTO}'s \texttt{rfifind} \citep{Ransom2001} with two second integration times. The resulting masks were applied to the data for subsequent processing. We incoherently dedispersed the data at $1000$ trial DMs ranging up to $\rm DM = 5000 \ \rm pc \ cm^{-3}$ with a step size of $5$. This is significantly higher than the inferred DM of $100 \ \rm pc \ cm^{-3}$ at the distance of PTF10hgi \citep{Deng2014}. Following dedispersion, we performed a standard red noise removal to properly normalize the time series. We searched individual scans for FRBs using the matched-filtering algorithm \texttt{single\_pulse\_search.py} \citep{Ransom2001}. No pulses are detected in the 40.5 min of on-source time.

%The redshift of PTF10hgi implies a dispersion measure from the intergalactic medium of $100 \ \rm pc \ cm^{-3}$ \citep{Deng2014}. Given a Milky Way contribution of $\sim 80 \ \rm pc \ cm^{-3}$ along the line of sight, we therefore incoherently dedispered the VLA data at $1000$ trial DMs ranging up to $\rm DM = 500 \ \rm pc \ cm^{-3}$ with a step size of $0.5$. We note that variances in the mapping from DM to redshift and the host contribution may yield a higher DM, however, our choice of maximum DM is comparable to that of \repeater{} ($\rm DM = 557 \ \rm pc \ cm^{-3}$) which is located at twice the distance. 

Following \citet{Cordes2003}, the minimum detectable flux density for an FRB above some S/N threshold is given by:
\begin{equation}
S_{\rm min} = \dfrac{\rm (S/N)_{\rm min}SEFD}{\sqrt{n_{\rm pol} \Delta \nu W}}
\end{equation}
where $n_{\rm pol}$ is the number of summed polarizations, $\Delta \nu$ is the bandwidth, $W$ is the intrinsic pulse width, and SEFD refers to the system equivalent flux density. We impose a signal-to-noise threshold of $10$ for a detection. Assuming a phasing efficiency factor of $0.9$ and a nominal $10$ ms pulse width, we find a minimum detectable flux density of $S_{\rm min}\approx 22$ mJy for our observations. 

We estimate an expected rate of FRBs with flux densities of $\gtrsim 22$ mJy assuming a universal luminosity function based on \repeater{} \ \citep{Nicholl2017c}. We find an expectation of $\approx 22$ FRBs per day, or $\approx 0.6$ per 40 min.  We further note that \repeater{} is known to undergo quiescent periods in which no FRBs are detected \citep{Chatterjee2017}.  Thus, our non-detection of FRBs from the location of PTF10hgi is not constraining at present.

\subsection{ALMA Observations}

We obtained millimeter observations with the Atacama Large Millimeter/submillimeter Array (ALMA) in Band 3 ($\sim 100$ GHz) on 2018 January 11 with a total on-source integration time of 22.2 minutes. Here we report results using the ALMA data products which utilize standard imaging techniques within CASA. The field is imaged using 2400 pixels and an image scale of 0.03 arcsec per pixel, MFS, Briggs weighting \citep{Briggs1995} with a robust parameter of 0.5, and a standard gridding convolution function. We do not detect  emission at the position of the VLA source, with a $3\sigma$ limit of $F_\nu(100\,{\rm GHz})\lesssim 44 \ \rm \mu$Jy. This indicates a radio to mm spectral index of $\alpha\lesssim 0$.

\subsection{{\it Hubble Space Telescope} Observations}

We observed the host galaxy of PTF10hgi on 2018 September 22 UT with the \textit{Hubble Space Telescope} (\textit{HST}) as part of program GO-15140 (PI: Lunnan), using the UVIS channel of the Wide Field Camera for Surveys 3 (WFC3). The galaxy was imaged in the F336W filter (corresponding to a rest-frame wavelength of $3055$ \AA, at the redshift of PTF10hgi) for two orbits, split into four dithered exposures for a total exposure time of 5570 s. We processed and combined the individual CTE-corrected images using the {Astrodrizzle} program from the {\tt Drizzlepac} software package provided by STScI\footnote{http://drizzlepac.stsci.edu/}, using a final {\tt pixscale} of 0.02'' per pixel and a {\tt pixfrac} value of 0.8. We show the resulting image in Figure~\ref{fig:image}.

\subsection{Astrometry}
\label{sec:astrom}

To determine the location of the radio source relative to the position of PTF10hgi and its host galaxy we first determine an astrometric solution for a wide-field $g$-band image centered on the host galaxy from the Inamori Magellan Areal Camera and Spectrograph (IMACS) on the Magellan Baade 6.5-m telescope using \textit{Gaia} sources from the latest data release.  We then register the smaller field of view \textit{HST} image on the \textit{Gaia} astrometric system using the IMACS image.  The resulting uncertainty in the astrometric tie between \textit{HST} and \textit{Gaia} is $\sigma_{\rm Gaia-host}=0.04''$. We find that the host galaxy is resolved into a bright central core, with diffuse extended emission and possibly other fainter emission knots (Figure~\ref{fig:image}).  The bright core is located at R.A.=${\rm 16^h37^m47^s.065}$, decl.=$+06^\circ12'31''.88$ (J2000), with a centroid uncertainty of $\sigma_{\rm host}=0.01''$.  

To determine the location of PTF10hgi in the same astrometric system we perform relative astrometry between the IMACS image and archival images of PTF10hgi from the Liverpool Telescope \citep{Inserra2013}, leading to a relative astrometric tie uncertainty of $\sigma_{\rm host-SN}=0.04''$.  The resulting absolute position of the SN (in the Gaia astrometric frame) is R.A.=${\rm 16^h37^m47^s.064}$, decl.=$+06^\circ12'31''.89$ (J2000), with a centroid uncertainty of $\sigma_{\rm SN}=0.02''$. Thus, the combined uncertainty in the absolute position of PTF10hgi is $0.05''$.

Comparing to the radio source position (\S\ref{sec:vla}) we conclude that the radio source is coincident with the optical position of PTF10hgi, with a nominal offset of $0.10''$ and a combined total uncertainty of $0.20''$ (dominated by the radio source positional uncertainty).  Furthermore, both the SN and the radio source are located near the core of the galaxy identified in the \textit{HST} image, with offsets of $0.02''$ ($\sigma=0.05''$) and $0.09''$ ($\sigma=0.20''$) for the optical SN and radio source, respectively. We note that the Gaia Celestial Reference Frame is consistent with the International Celestial Reference Frame to within 0.5 milliarcseconds, and thus the uncertainty in the astrometric tie between the radio and optical images is negligible relative to the positional uncertainty of the radio source.

\section{Origin of the Radio Emission}
\label{sec:origin}

\begin{figure*}
\includegraphics[width=\textwidth]{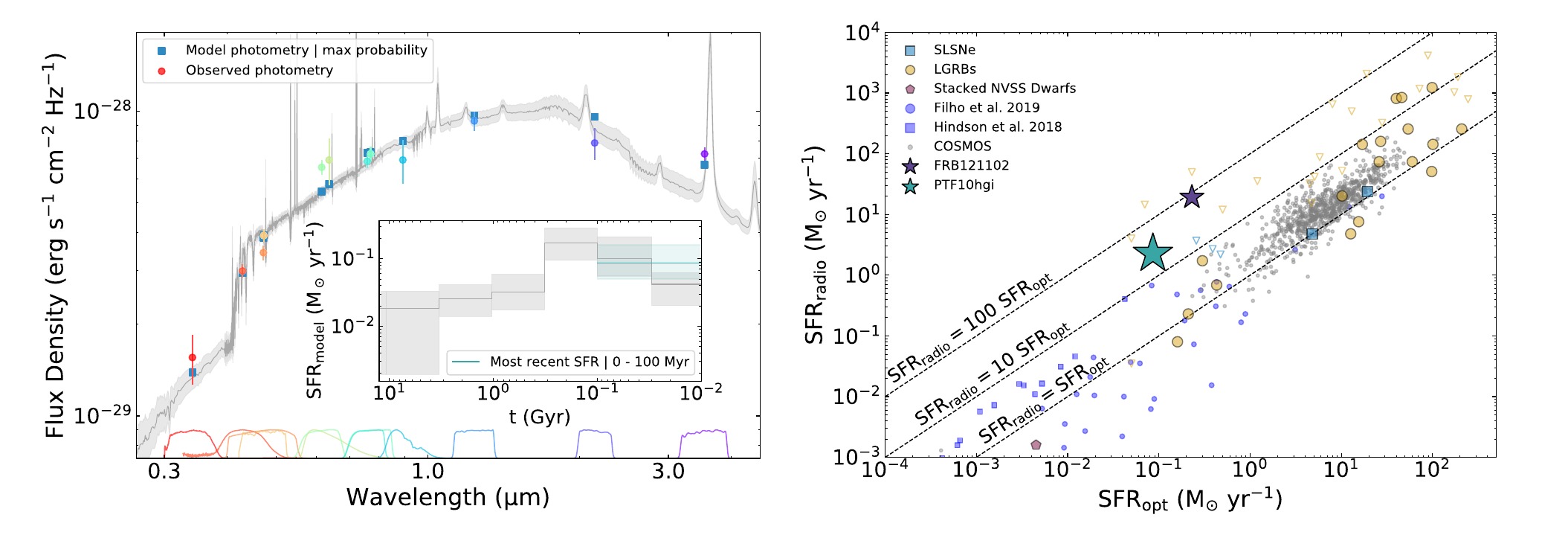}
\caption{\textit{Left:} UV to NIR SED of the host galaxy of PTF10hgi (color points), along with the best-fit model photometry from \texttt{Prospector} (blue points) and the 16th and 84th percentile range of the model SEDs (grey). The inset shows the 16th, 50th, and 84th percentiles of the marginalized star formation history, as well as the time-averaged SFR over the past 100 Myr, corresponding to the timescale of radio emission due to star formation. \textit{Right:} Radio versus optical SFRs for PTF10hgi (green star), \repeater{} (purple star; assuming a star formation origin for the radio emission; \citealt{Bassa2017}), LGRB hosts (yellow; \citealt{Perley2013a,Perley2015,Greiner2016}), and SLSN hosts (cyan; \citealt{Hatsukade2018}).  Upper limits are shown as open triangles.  We also show the results for nearby dwarf galaxies from a number of surveys (blue points; \citealt{Roychowdhury2012,Hindson2018,Filho2019}), as well as star forming galaxies at $z\lesssim 0.5$ from the VLA-COSMOS survey (grey points; \citealt{Smoli2017}). Dashed lines indicate ratios of $\rm SFR_{radio} = SFR_{opt}, 10 \times SFR_{opt}$, and $\rm 100 \times SFR_{opt}$.}
\label{fig:sfrs}
\end{figure*}

Given the spatial coincidence of the radio source and PTF10hgi (and its host galaxy) we use the redshift of $z=0.098$ to determine a radio source luminosity of $L_\nu(6\,{\rm GHz})=(1.1\pm 0.2)\times 10^{28}$ erg s$^{-1}$ Hz$^{-1}$ for a luminosity distance of 465 Mpc\footnote{Throughout the paper, we use the standard cosmological constants with $H_0 = 67.7 \ \rm km \ s^{-1} \ Mpc^{-1}$, $\Omega_m = 0.3$, $\Omega_{\lambda} = 0.7$.}.  With a single epoch and single frequency detection, and given the coincidence with both the SLSN position and the host galaxy center, the radio emission could result from several processes that we investigate below.  We show that an origin due to star formation activity, an active galactic nucleus (AGN), an off-axis relativistic jet, or a spherical non-relativistic outflow are all unlikely, and would be quite unusual. This leaves open the possibility that the radio emission instead shares a common origin with the quiescent source coincident with \repeater{}.  

\subsection{Star formation Activity}

The host of PTF10hgi is a low mass, low metallicity dwarf galaxy, with $M_B\approx -15.9$ mag ($\approx 0.017$ L$^*$), $M_*\approx 10^8$ M$_\odot$, and $12+{\rm log[O/H]}\approx 8.3$ \citep{Lunnan2014,Perley2016a,Schulze2018}. It has a relatively low star formation rate (SFR) of $\approx 0.01-0.04$ M$_{\odot}$ yr$^{-1}$ based on H$\alpha$ emission \citep{Lunnan2014,Leloudas2015,Perley2016a}) and $\approx 0.1-0.2$ M$_{\odot}$ yr$^{-1}$ based on modeling of the UV to NIR spectral energy distribution (SED; \citealt{Perley2016a,Schulze2018}).

To test a star formation activity origin for the radio emission, we calculate the radio-inferred SFR using the expression from \citet{Greiner2016}, which is extrapolated from the 1.4 GHz radio luminosity SFR relation of \citet{Murphy2011b} assuming a power law $F_\nu \propto \nu^{\alpha}$ and accounting for proper $k$-corrections:
\begin{equation}
{\rm SFR_{radio}} = 0.059\, {\rm M_{\odot} \ yr^{-1}}\, F_{\rm \nu,\mu Jy}d_{L,{\rm Gpc}}^2\nu_{\rm GHz}^{-\alpha}(1+z)^{-(\alpha + 1)},
\end{equation}
where $F_\nu$ is the observed flux density at a frequency $\nu$, $d_L$ is the luminosity distance at a redshift $z$ (465 Mpc for PTF10hgi), and here we adopt a canonical value of $\alpha = -0.75$ (e.g., \citealt{Condon1992,Tabatabaei2017}). We find a radio-inferred SFR of $2.3\pm 0.3$ M$_{\odot}$ yr$^{-1}$. This is a factor of $\approx 12-230$ times higher than the SFR based on H$\alpha$ and SED modeling. 
Given the range of quoted SFR values from the literature, we independently model the host SED using the \texttt{Prospector} software package \citep{Leja2017}; we use the magnitudes reported by \citealt{Lunnan2014} and \citealt{Perley2016a}. The model accounts for dust attenuation and emission by imposing a two component dust screen and energy balance (i.e.,  that stellar emission absorbed by dust is re-radiated at far-IR wavelengths). The net effect is that the inferred SFR accounts for dust obscuration of both young stars within molecular clouds and HII regions, as well as stellar and nebular emission due to a diffuse dust screen. 

The resulting SED and star formation history are shown in Figure~\ref{fig:sfrs}. We find peak star formation activity about $0.1-0.3$ Gyr ago (with $0.2$ M$_\odot$ yr$^{-1}$), with a steady decline since, and a present-day ($\lesssim 30$ Myr) SFR of $\approx 0.04$ M$_\odot$ yr$^{-1}$ (in agreement with the H$\alpha$ values). For the purpose of comparison to the radio-inferred SFR, we average the star formation history over the past $0.1$ Gyr, corresponding to the timescale over which supernovae-accelerated electrons radiate their energy via radio synchrotron emission \citep{Condon1992} and find ${\rm SFR_{opt}=0.09 \  \rm M_{\odot} \ yr^{-1}}$.  This indicates that if the radio emission is due to star formation activity, then ${\rm SFR_{radio}/SFR_{opt}}\approx 26$ (i.e., about $96\%$ of the star formation activity is completely dust obscured).

Such a high ratio of obscured star formation activity is typical of LIRGs and ULIRGs, but is not expected for low mass and low metallicity galaxies such as the host of PTF10hgi; from our SED modeling, we infer a stellar mass of $3.1^{+1.4}_{-1.6}\times 10^{8} \ \rm M_{\odot}$. In Figure~\ref{fig:sfrs} we compare the radio versus optical SFRs for the host of PTF10hgi to those of previous SLSN and LGRB hosts \citep{Perley2013a,Greiner2016,Perley2015,Hatsukade2018}, as well as to samples of dwarf galaxies from a number of surveys \citep{Roychowdhury2012,Hindson2018,Filho2019}, and star forming galaxies at $z\lesssim 0.5$ from the COSMOS survey \citep{Smoli2017}.  We find that SLSN and LGRB hosts span values of ${\rm SFR_{radio} / SFR_{opt}}\approx 1-10$, with only the most prodigiously star forming hosts (${\rm SFR_{opt}}\gtrsim 10$ M$_\odot$ yr$^{-1}$) approaching the upper end of ${\rm SFR_{radio} / SFR_{opt}}\approx 10$.  For the COSMOS sample the mean and standard deviation are ${\rm SFR_{radio} / SFR_{opt}}\approx 2.2\pm 1.2$, more than an order of magnitude below the value for PTF10hgi. Similarly, for dwarf galaxies with low optical star formation rates comparable to the host of PTF10hgi, the ratios span ${\rm SFR_{radio} / SFR_{opt}}\approx 0.1-10$, with a typical value of $\approx 1$. Thus, we consider a star formation origin for the radio emission to be unlikely, but stress that if this was indeed the case, then the host of PTF10hgi would represent quite an unusual galaxy. 

Instead, we note that the large radio luminosity in comparison to the expected contribution from star formation activity is reminiscent of \repeater{} and its host galaxy, with $L_\nu(6\,{\rm GHz})\approx 2.2\times 10^{29}$ erg s$^{-1}$ Hz$^{-1}$ \citep{Chatterjee2017} and ${\rm SFR_{radio} / SFR_{opt}}\approx 84$ if the radio emission is interpreted as being due to star formation (Figure~\ref{fig:sfrs}).

Our ALMA non-detection at 100 GHz does not provide meaningful constraints on a star formation origin since at that frequency synchrotron emission still dominates, with an expected  $\alpha\approx -0.75$ (compared to our shallow limit of $\alpha\lesssim 0$).  On the other hand, observations at frequencies of several hundred GHz can directly probe the presence of dust continuum emission and therefore provide an independent measure of obscured star formation. For example, we expect a flux density of $\approx 0.4$ mJy at $400$ GHz if the host indeed has an obscured SFR of 2.3 M$_\odot$ yr$^{-1}$; a non-detection well below this value, which can be obtained with ALMA in $\approx 1.5$ hours, will definitively rule out obscured star formation as the origin of the radio emission. Similarly, high angular resolution observations with the VLBA can rule out a star formation origin if they show that the radio emission is unresolved at a parsec-scale.

\subsection{Active Galactic Nucleus}

Based on the proximity of the radio source to the optical center of the host galaxy, we investigate an AGN origin. The host galaxy shows no evidence for an AGN from optical emission lines, and instead resides well within the star forming branch of the BPT diagram \citep{Lunnan2014,Leloudas2015,Perley2016a}. Nevertheless, we place limits on a putative black hole mass assuming an AGN origin and using the ``fundamental plane'' of black hole activity \citep{Merloni2003}. Given the radio luminosity and a \textit{Swift}/XRT upper limit of $L_X\lesssim 4\times 10^{42}$ erg s$^{-1}$ \citep{Margutti2018} we find a lower limit for the mass of the black hole of $\gtrsim 1.4 \times 10^7 \ \rm M_{\odot}$. This value is unexpectedly large, $\gtrsim 0.05$ of the galaxy's stellar mass, while black hole masses in dwarf galaxies are generally $\lesssim 10^{-3}$ of the stellar mass \citep{Reines2013}. 

Conversely, the lack of X-ray emission and the absence of AGN signatures in the optical spectrum could be consistent with a low-luminosity, radio-loud AGN \citep{Mauch2007}, as has been suggested for the persistent radio source coincident with \repeater{} \citep{Tendulkar2017,Marcote2017}. Indeed, five such LLAGN were recently discovered by their radio emission \citep{Park2016}, however, these galaxies have much larger stellar masses ($\sim 10^{10} \ \rm M_{\odot}$) relative to the host of PTF10hgi ($\sim 10^8 \ \rm M_{\odot}$). Furthermore, the prevalence of AGN in dwarf galaxies is extremely low; for example, a search for AGN in dwarf galaxies ($10^{8.5}-10^{9.5} \ \rm M_{\odot}$) based on pre-selection using optical emission lines yielded a detection rate of $\lesssim 1\%$ \citep{Reines2013}. To date, only two AGN in dwarf galaxies have been found to host nuclear radio sources \citep{Reines2011,Reines2014}.

Thus, we consider the AGN scenario to be unlikely, but note that if this was shown to be the case it would represent quite a rare discovery, especially given that the host galaxy was ``selected'' for the occurrence of a SLSN, which itself should not be correlated with AGN activity. Nevertheless, if FRBs were found to occur preferentially near radio sources associated with AGN, it may suggest that SLSNe require special environments (e.g., near massive black holes) to produce FRB emission \citep{Michilli2018}.

\begin{figure}
\includegraphics[width=1.1\columnwidth]{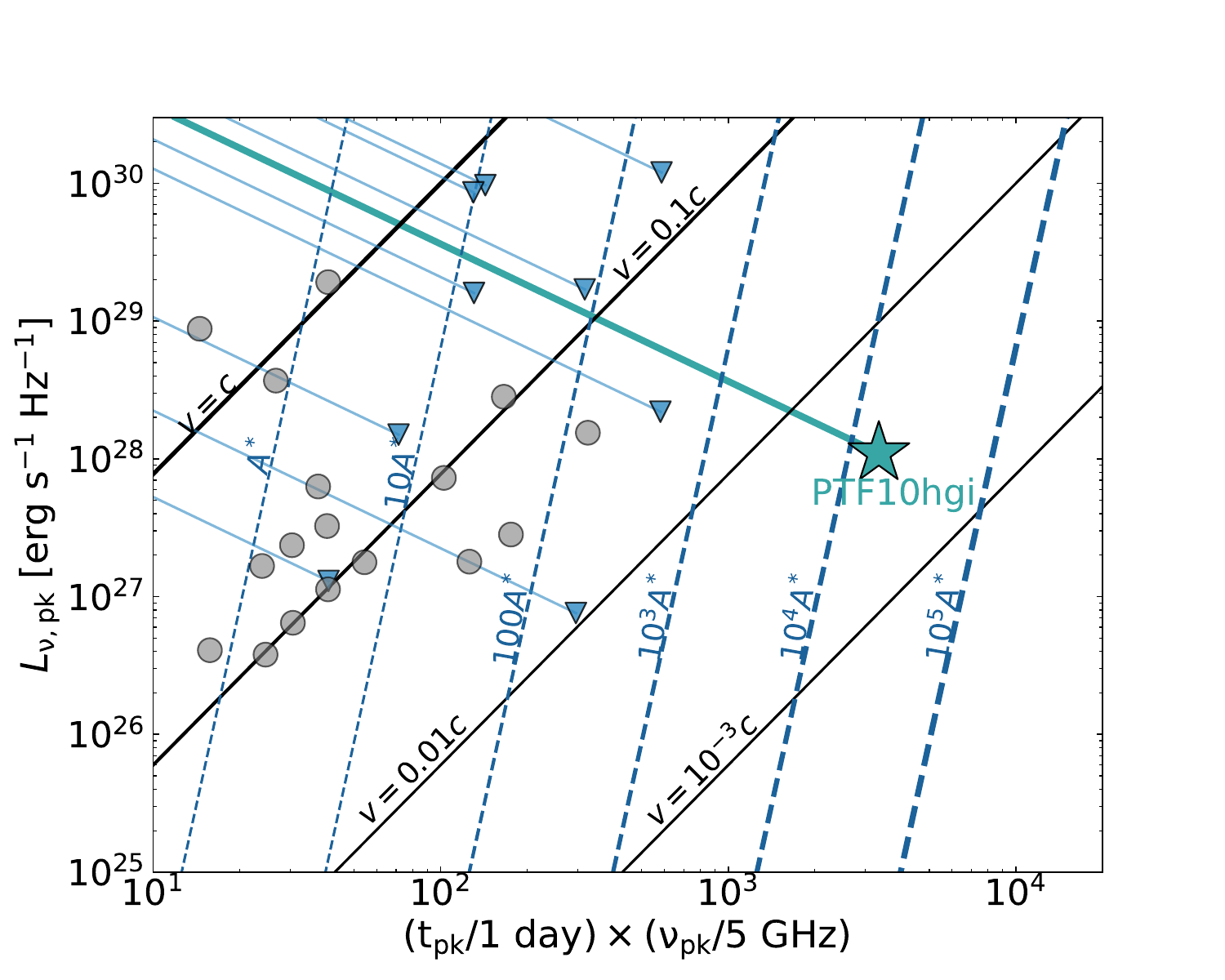}
\caption{Peak radio luminosity ($L_{\rm \nu,pk}$) versus the product of peak  frequency and time ($\nu_{\rm pk}\times t_{\rm pk}$). Black and blue lines correspond to constant shock velocity and mass-loss rate, respectively, following the prescription for self-absorbed synchrotron emission from a non-relativistic spherical blastwave \citep{Chevalier1998}, with $\epsilon_e = \epsilon_B = 0.1$.  The mass-loss rate is parameterized in terms of the wind mass-loss parameter $A^*$ (equal to 1 for $\dot{M}=10^{-5} \rm \ M_{\odot} \ yr^{-1}$ and a wind velocity of $10^3 \ \rm km \ s^{-1}$). We show the radio detection of PTF10hgi assuming that our observation corresponds to the peak luminosity at 6 GHz (star), as well as an extension to earlier peak times (line; $L_{\rm \nu,pk}\propto t^{-1}$). Also shown are the data for Type Ib/c SNe, including those associated with nearby LGRBs \citep{Soderberg2005,Margutti2018b}, and upper limits for SLSNe from \citet{Coppejans2018}, with individual lines for each source accounting for a possible peak at earlier times.}
\label{fig:blastwave}
\end{figure}

\subsection{External Blastwave}

We next consider whether the radio emission could be due to external shock interaction between outflowing ejecta and the circumstellar medium (CSM). Such emission may arise from an initially off-axis relativistic jet that has decelerated and spread into our line of sight at late time \citep{Rhoads1997,Sari1999}, or from the fastest layers of the (quasi)-spherical SN ejecta, as observed in stripped-envelope Type Ib/c SNe \citep{Chevalier1998}. In both scenarios we can use the observed radio emission to estimate the properties of the outflow and CSM, and hence to assess the feasibility of this explanation by comparing to existing observations of LGRBs, SLSNe, and Type Ib/c SNe.s\begin{figure*}
\includegraphics[width=\textwidth]{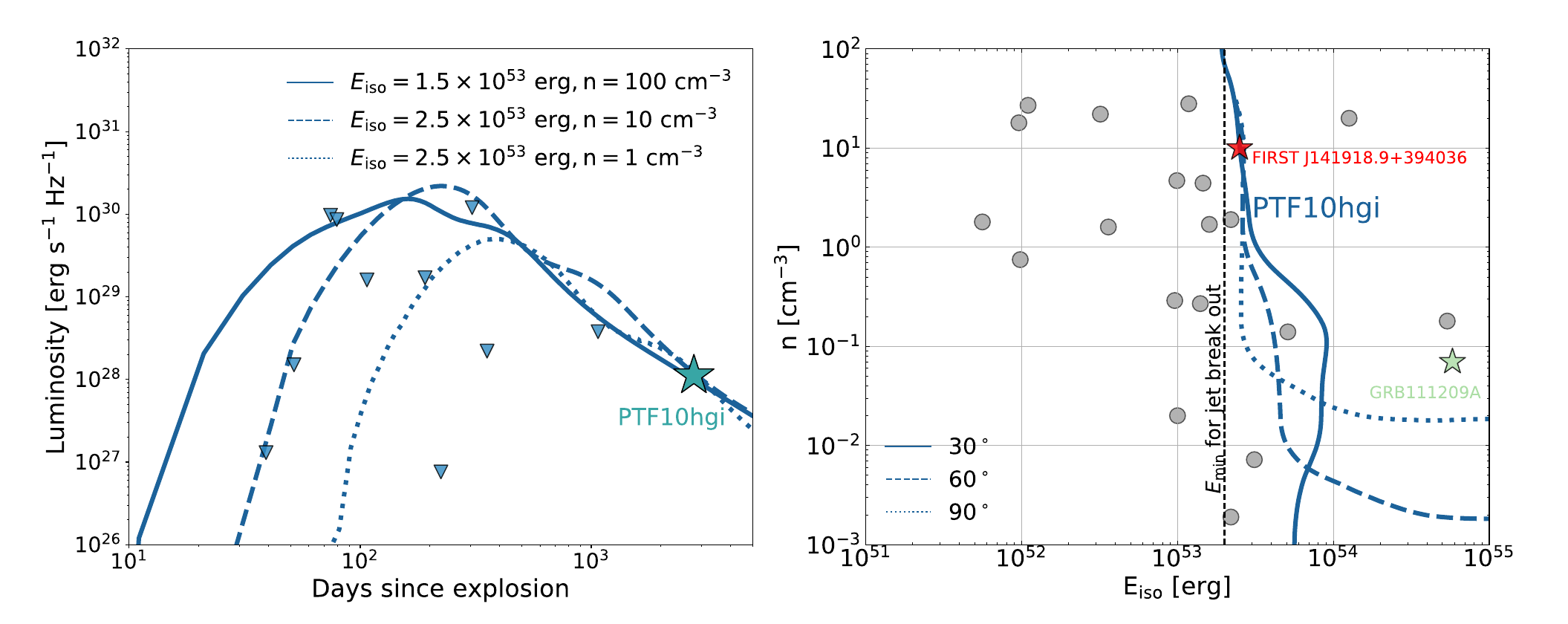}
\caption{\textit{Left:} Representative off-axis jet light curves at 6 GHz for $\theta_{\rm obs} = 60^{\circ}$ and a range of jet energies and CSM densities that are consistent with the radio detection of PTF10hgi. For comparison, we also plot upper limits for other SLSNe from \citet{Coppejans2018}, as well as the limit for SN2015bn at $\delta t \approx 1070$ d from \citet{Nicholl2018b} converted to 6 GHz assuming a typical spectral index of $-0.7$. \textit{Right:} Constraints on the jet energy and CSM density for an off-axis jet assuming a jet opening angle $\theta_j = 10^\circ$ and viewing angles of $\theta_{\rm obs} = 30^\circ$ (solid), $60^\circ$ (dashed), and $90^\circ$ (dotted). Individual curves trace out the allowed parameter space for an off-axis jet based on the 6 GHz radio detection. The vertical line at $E_{\rm iso} = 2\times 10^{53}$ erg marks the minimum required energy for a successful jet to break through the SN ejecta, based on the inferred properties of PTF10hgi \citep{Nicholl2017b,Duffell2018,Margalit2018c}. For comparison, we also show the results for FIRST J141918.9+394036 from \citet{Law2018} and the ultra-long GRB 111209A \citep{Stratta2013} as well as LGRBs from the literature \citep{Berger2001,Panaitescu2002,Berger2003,Yost2003,Chevalier2004,Chandra2008,Cenko2010,Laskar2015}.}
\label{fig:jets}
\end{figure*}

\subsubsection{Supernova Ejecta}

We first investigate the scenario of radio emission from the spherical SN ejecta. In Figure~\ref{fig:blastwave} we show the radio detection in the phase-space of peak luminosity versus peak time assuming that the observed emission corresponds to the peak of the radio SED at the time of the observation (for comparison with Type Ib/c SNe from the literature we also make the standard assumption of $\epsilon_e = \epsilon_B = 0.1$, where $\epsilon_e$ and $\epsilon_B$ refer to the fraction of post-shock energy in electrons and the magnetic field, respectively). From this we infer a low ejecta velocity of $v_{\rm ej}\approx 10^3$ km s$^{-1}$ and a dense CSM with a wind parameter of $A\equiv\dot{M}/4\pi v_w\approx 2\times 10^4 A^*$ (where $A^*$ is the wind mass-loss parameter and is equal to $1$ for $\dot{M}=10^{-5} \ \rm M_{\odot} \ yr^{-1}$ and a wind velocity of $10^{3} \ \rm km \ s^{-1}$), or a progenitor mass loss rate of $\dot{M}\approx 0.2$ M$_\odot$ yr$^{-1}$ for $v_w=1000$ km s$^{-1}$.   

These values are in stark contrast to radio-emitting Type Ib/c SNe for which the inferred values are $v_{\rm ej}\sim 0.1$c and $A\sim 1-100 \ A^*$ (e.g., \citealt{Berger2002,Soderberg2005,Soderberg2012}).  However, in the context of this scenario the actual peak time at 6 GHz may have occurred earlier than our observation (with a correspondingly higher peak luminosity), and we therefore extrapolate the observed emission as a power law given by $L_{\nu, p}\propto t^{-1}$ (e.g., \citealt{Berger2002}). With this extrapolation the radio emission from PTF10hgi would have been more luminous in the radio than any known Type Ib/c SN, including relativistic events such as SN1998bw, if it had peaked on the typical range of timescales (Figure~\ref{fig:blastwave}), or equivalently it would require an unusually high mass loss rate for the typical range of inferred ejecta velocities. Similarly, existing limits for SLSNe show no evidence for outflows comparable to Type Ib/c SNe. We therefore consider this scenario unlikely, but note that future observations to search for power law fading of the source will further test this possibility.

While the persistent radio source associated with \repeater{} cannot be placed in Figure~\ref{fig:blastwave} due to its unknown age, the fact that it is likely older than PTF10hgi and that its luminosity is higher by an order of magnitude would make it even more anomalous compared to the Type Ib/c SN sample, and thus similar to PTF10hgi in this regard.

\subsubsection{Off-Axis Jet}

In the context of an off-axis jet origin for the radio emission, we constrain the required combination of jet energy and CSM density by generating a grid of afterglow models for viewing angles of $30^\circ$, $60^{\circ}$, and $90^{\circ}$ using the two-dimensional relativistic hydrodynamical code {\tt Boxfit v2} \citep{vanEerten2012}. We assume a CSM with constant density ($n$), a jet opening angle of $10^\circ$, and microphysical parameters of $\epsilon_e=0.1$, $\epsilon_B=0.01$, and $p=2.5$, typical of LGRBs (e.g., \citealt{Curran2010,Laskar2013,Wang2015,Laskar2016,Alexander2017}). The results are summarized in Figure~\ref{fig:jets}.  We find that the observed flux density can be reproduced for an isotropic equivalent jet energy $E_{\rm iso} \sim (3-5)\times 10^{53}$ erg and a wide range of CSM densities ($n\sim 10^{-3} - 10^2 \ \rm cm^{-3}$, depending on the viewing angle).  The corresponding beaming corrected energy is $\sim (5-8)\times 10^{51}$ erg.

Previous radio searches for off-axis jets in SLSNe have yielded only non-detections (e.g., \citealt{Coppejans2018,Nicholl2018b}), ruling out the presence of jets with an energy scale similar to the one required for PTF10hgi (Figure~\ref{fig:jets}). Even relative to the sample of LGRBs, an off-axis jet in PTF10hgi would be among the most energetic observed to date (Figure~\ref{fig:jets}), although we note that the large inferred energies are consistent with the ultra-long GRB 111209A \citep{Stratta2013}, which has been argued to be associated with the supernova SN2011kl \citep{Greiner2015}. Thus, based on the lack of previous evidence for similarly powerful jets in SLSNe, and the large inferred energy relative to most LGRBs, we conclude that an off-axis jet would be unusual. Nevertheless, it would be the first evidence for such an outflow in a SLSN if this was indeed the case, and would directly implicate a central engine as the energy source of the explosion.

On the other hand, we note that for the explosion parameters of PTF10hgi \citep{Nicholl2017b}, the analysis of \citet{Margalit2018c} indicates that for a $10^\circ$ jet to break out of the SN ejecta requires a minimal energy of $E_{\rm iso}\gtrsim 2\times 10^{53}$ erg.  That the inferred energy of the jet from our analysis above is a factor of $2-3$ times higher than this threshold value indicates at least a self-consistency to the jet scenario. Furthermore, the allowed jet energies and CSM densities are consistent with the inferred afterglow parameters for the extragalactic transient FIRST J141918.9+394036 ($E_{\rm iso} = 2\times 10^{53}$ erg and $n=10 \ \rm cm^{-3}$) which is also located in a dwarf galaxy  \citep{Law2018}. The discovery of jetted emission from PTF10hgi would thus favor the hypothesis that the observed radio emission from FIRST J141918.9+394036 is due to a SLSN. 

While we note that the persistent radio source associated with \repeater{} cannot be placed in Figure~\ref{fig:jets} due to its unknown age, the fact that it is likely older than PTF10hgi and that its luminosity is higher by an order of magnitude would make it even more anomalous compared to the LGRB sample, and thus similar to PTF10hgi in this regard. Furthermore, in the scenario of an off-axis jet for FRB121102, the shocked ISM plasma would not produce the large rotation measure (RM) observed in the bursts themselves, thereby requiring that the RM-producing medium is separate from that generating the persistent source.

Further multi-frequency radio observations to constrain the SED, which is expected to be optically thin, and to search for fading will test this scenario (Figure~\ref{fig:nebula}).

\subsection{Magnetar Nebula}
\label{sec:engine}

Here we explore the possibility that the observed radio emission is due to a pulsar wind nebula powered by a young magnetar embedded in the SN ejecta \citep{Metzger2014,Metzger2017,Omand2017,Margalit2018a}. In this framework, radio emission is expected from SLSNe on $\sim$decade timescales, as the ejecta expand and become transparent to free-free absorption at GHz frequencies. Indeed, such a nebula has been proposed as the origin of the persistent radio source associated with \repeater{} \citep{Kashiyama2017,Metzger2017,Margalit2018a,Margalit2018b}.

Following the prescription of \citet{Margalit2018a}, we compute the time-dependent evolution of the ionization structure of the ejecta for PTF10hgi using the photoionization code \texttt{CLOUDY} \citep{Ferland2013}. Specifically, we assume photo-ionization by a magnetar engine to constrain the free-free transparency timescale $t_{\rm ff}$, where the free-free optical depth scales as $\tau_{\rm ff} \sim t^{-4.5}$ \citep{Margalit2018a}. We use the ejecta and engine properties inferred from a model fit to the light curves of PTF10hgi \citep{Nicholl2017b}, namely a spin of $P = 4.8$ ms, a magnetic field of $B = 2\times 10^{14}$ G, an ejecta mass of $M_{\rm ej} = 2.2 \ \rm M_{\odot}$, and an ejecta velocity of $v_{\rm ej} = 5.1 \times 10^3 \ \rm km \ s^{-1}$. These parameters are fully within the distribution of the SLSN sample, with the mass and velocity representative of the low end, and the magnetic field and spin period corresponding to the large end. Assuming in addition a power-law energy injection rate into the nebula ($L\propto t^{-2}$), we find that $t_{\rm ff}\approx 4.8$ and $1.4$ years at 6 and 100 GHz, respectively, consistent with our radio detection at about 7.5 years post-explosion.

\begin{figure*}
\includegraphics[width=\textwidth]{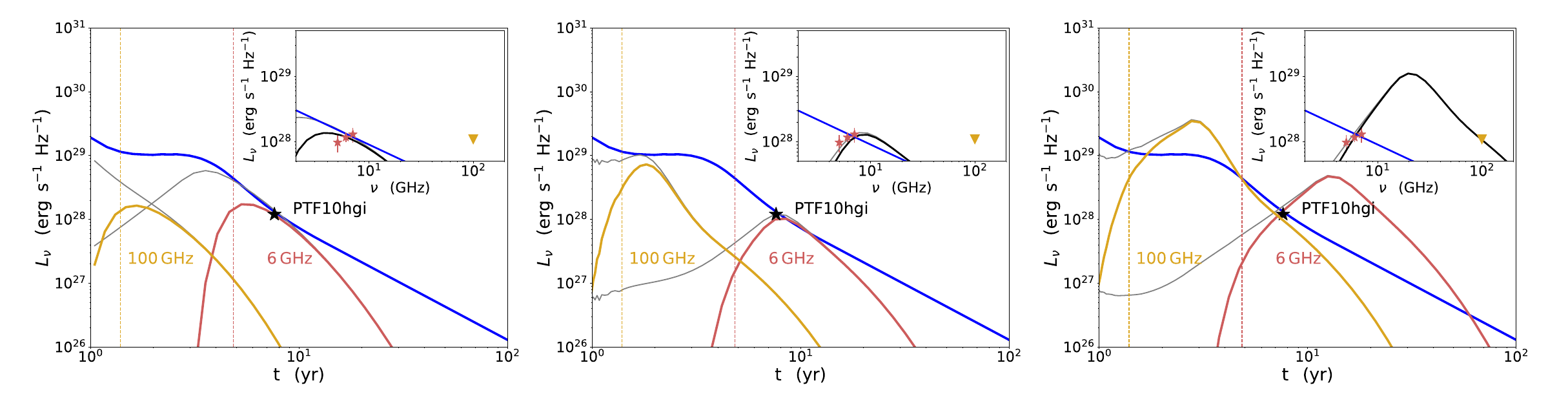}
\caption{Nebula models for the radio source associated with PTF10hgi based on the prescription for \repeater{} from \citet{Margalit2018b}. We show model light curves at 6 GHz (red) and 100 GHz (yellow) in the main panels and SEDs in the insets, in comparison to the data, for three cases.  From left to right, the first model is identical to that for \repeater{} with the magnetic energy scaled down by a factor of $\approx 20$ (i.e., $E_{\rm B_*} = 2.3\times 10^{49}$ erg); this model leads to an optically thin SED at $\gtrsim 4$ GHz. The middle and right panels correspond to limiting cases for a synchrotron self-absorbed nebula that accommodate both the 6 GHz detection and the 100 GHz limit (see \S\ref{sec:engine} for details). The vertical dashed lines indicate the free-free transparency times ($\tau_{\rm ff} = 1$) and the grey curves show the unabsorbed light curves.  For the purpose of comparison, the blue curve depicts a representative off-axis jet model with $\theta_{\rm obs} = 30^\circ$, $E_{\rm iso} = 2.5\times 10^{53}$ erg, and $n=10 \rm \ cm^{-3}$ (light curve in the main panels and SED in the insets).}
\label{fig:nebula}
\end{figure*}

In Figure~\ref{fig:nebula}, we plot three representative nebula models for PTF10hgi based on the inferred model for \repeater{} from \citet{Margalit2018b} in which the quiescent radio emission is due to a magnetized ion-electron wind nebula. This model is motivated by the observed rotation measure for \repeater{} and its time derivative \citep{Michilli2018}, as well as the persistent source luminosity and spectrum.  Given the single epoch observation of PTF10hgi, we modify the best fit model parameters for \repeater{} to fit the observed luminosity and upper limit at 6 and 100 GHz, respectively. The model parameters include the magnetic energy of the magnetar ($E_{\rm B_*}$), the nebula velocity ($v_{\rm n}$), the onset of the active period ($t_0$), the power law index describing the rate of energy input into the nebula ($\alpha$), the magnetization of the outflow ($\sigma$), and the mean energy per particle ($\chi$). We fix $\sigma = 0.1$ and $\chi = 0.2$ GeV as in the case of \repeater{}. For the first model in Figure~\ref{fig:nebula}, the inferred parameters are identical to ``model A'' for \repeater{} from \citet{Margalit2018b} with $t_0 = 0.2$ years, $v_n = 3\times 10^8 \ \rm cm \ s^{-1}$, $\alpha = 1.3$, and the magnetic energy scaled down by a factor of $\approx 20$ to $E_{\rm B_*} = 2.3 \times 10^{49}$ erg. This directly scaled model can adequately explain the observed radio emission, and predicts an optically thin spectrum in our observing band, consistent with the inferred range of values from the VLA data ($-0.8$ to $+2.5$; \S\ref{sec:vla}).

We also explore models in which the emission at 6 GHz is marginally or fully synchrotron self-absorbed (SSA), with the latter model constrained by the non-detection at 100 GHz (Figure~\ref{fig:nebula}).  We constrain the allowed model parameters under the assumption that the magnetic field in the nebula is given by $B \sim (\sigma \dot{E} t/R^3)^{1/2}$, corresponding to a luminosity $L_\nu (\nu \ll \nu_{\rm ssa} )\sim R^{11/4} (\sigma \dot{E} t )^{-1/4}$. Thus, for a fixed time $t$ and observed luminosity $L_\nu ( \nu \ll \nu_{\rm ssa})$, we can constrain the model parameters by satisfying $R \sim \dot{E}^{1/11}$ and further requiring that the spectrum does not overproduce the non-detection at 100 GHz and that the self absorption frequency $\nu_{\rm ssa}>6$ GHz. This allows for an upper and lower limit on the allowed values of $\dot{E}$ and $R$, corresponding to the two limiting cases shown in Figure~\ref{fig:nebula}.  We find that the relevant physical parameters are $\dot{E} \sim 3\times 10^{40} \rm \ erg \ s^{-1}$ and $R \sim 2\times 10^{16} \ \rm cm$ in the first scenario and $\dot{E} \sim 3\times 10^{39} \ \rm erg \ s^{-1}$ and $R \sim 1.7\times 10^{16} \rm cm$ in the second scenario.  The inferred source size corresponds to a velocity of about $850$ km s$^{-1}$, which is slower than the ejecta velocity of PTF10hgi ($v_{\rm ej} = 5.1 \times 10^3 \ \rm km \ s^{-1}$) and is thus consistent with a nebula expanding within the SN ejecta.

We therefore conclude that the model of a central engine driven nebula is fully consistent with the observations, both in terms of the free-free transparency timescale and in terms of explaining the source luminosity and SED with a reasonable range of parameters.

This model can be further tested in several ways.  First, additional observations covering frequencies of $1-40$ GHz will establish the shape of the SED and the location of $\nu_{\rm ssa}$; this is the only model that can account for a self-absorbed SED at $\gtrsim {\rm few}$ GHz. Second, continued temporal coverage will determine whether the source is rapidly fading or rising, both of which are in contrast to the expectations of an off-axis jet (Figure~\ref{fig:nebula}).  Third, the predicted angular size of the nebula is $\sim 10$ $\mu$as, and therefore strong refractive scintillation with a flux density modulation of tens of percent is expected.  This is in direct contrast to an off-axis jet, with an angular size of $\sim{\rm mas}$ for which no scintillation is expected. A modest time investment of several hours of High Sensitivity Array VLBI observations would be sufficient to detect a point source at the $5 \sigma$ level. Conversely, marginally resolved emission would point to the presence of an off-axis jet with an inferred angular size of $\sim$ mas.

\subsection{A Continuously Bursting Source}

Finally, we briefly consider the speculative possibility that the observed emission is due to a continuously bursting source, with bursts occurring rapidly enough to produce a quasi-steady source during our VLA observation, but with no flares bright enough to be detected in our phased-array data. The mean flux density for aperiodic bursts of width $W$ emitted with a constant rate $\eta$ and mean amplitude $\langle a\rangle$ is given by $\langle S \rangle = \eta \langle a\rangle W$. In the limit of Poisson statistics, the burst duty cycle $\eta W$ can be used to place a limit on the minimum mean amplitude by requiring $\eta W \lesssim 0.5$, i.e., that pulses are emitted roughly half of the time as the beam of emission rotates into the line of sight (in analogy with pulsars). This implies $\langle a\rangle \gtrsim 2 \langle S \rangle \sim 0.1$ mJy for PTF10hgi, or a factor of about 220 times below the sensitivity of our phased-array VLA search (\S\ref{sec:phased}). 

The lack of bursts detected from PTF10hgi in 40 minutes of VLA phased-array observations implies that the average burst amplitude is below the minimum detectable flux density of the observation, i.e., $\langle a \rangle < S_{\rm min}$. This in turn allows for a lower limit on the rate of bursts given by $\eta > \langle S \rangle/S_{\rm min} W$. For our limit of $S_{\rm min}\approx 22$ mJy and a typical burst width of $10$ ms, the source flux density of $\approx 50 \ \mu$Jy requires a burst rate of $\eta \gtrsim 0.2 \ \rm s^{-1}$. 

This rate is three orders of magnitude larger than for \repeater{}, for bursts of a similar luminosity ($\sim 2.5\times 10^{-4} \ \rm s^{-1}$; \citealt{Nicholl2017c}). This therefore suggests that individual bright bursts well above our limit of 22 mJy should have been detected. We therefore conclude that this scenario is unlikely, but future more sensitive searches for bursts with the GBT or Arecibo will further test this scenario.

\section{Conclusions and Future Observations}
\label{sec:conc}

We presented radio and mm observations of the SLSN PTF10hgi about 7.5 years post-explosion that reveal the presence of an unresolved radio source coincident with the SN (and host galaxy) position, with a luminosity of $L_\nu(6\,{\rm GHz})\approx 1.1\times 10^{28}$ erg s$^{-1}$ Hz$^{-1}$.  This is the first case of radio emission spatially coincident with a known SLSN. We explored multiple origins for the radio emission, including star formation activity, an AGN origin, emission due to an external blastwave (relativistic and non-relativistic), and emission from a compact central engine.   

If the observed radio emission is due to star formation activity, then the large ratio of $\rm SFR_{radio} / SFR_{opt}\approx 26$ implies that $96\%$ of the star formation in the host galaxy is completely obscured, typical of LIRGs and ULIRGs, but unprecedented for low mass, low metallicity galaxies. Indeed, this would represent the most highly dust-obscured SLSN (or LGRB) host galaxy observed to date. This scenario can be definitively tested using high-frequency ALMA observations to probe the presence of thermal dust emission, and with milliarcsecond resolution radio VLBI imaging to determine the angular extent of the emission region.

Alternatively, the radio emission may be due to a radio-loud AGN, but the lack of any other AGN signatures in the host galaxy, the low occurrence rate for nuclear radio sources in dwarf galaxies, and the high black hole mass implied by the fundamental plane of black hole activity all suggest that the presence of a radio-loud AGN would be quite unusual. Improved astrometry from radio VLBI observations can be used to test whether the radio source is offset from the host center, thereby further disfavoring an AGN. 

In the context of radio emission from the SN ejecta, we find that the timescale and luminosity of the observed radio emission imply an ejecta velocity and/or progenitor mass loss rate that are at least a few times larger than those in stripped-envelope Type Ib/c SNe. Similarly, if the radio emission is due to an off-axis jet, this would be one of the most powerful jets observed to date in comparison to LGRBs, and the first time that such a jet has been detected in a SLSN (despite previous searches).  However, we note that the implied jet energy is above the threshold for a jet to break out of the PTF10hgi ejecta. Similarly, the inferred jet energies and CSM densities are similar to that of FIRST J141918.9+394036, suggesting that both events may represent jetted emission from a SLSN. For both scenarios, continued radio observations to determine the spectral energy distribution and to search for fading will provide a powerful test.

Finally, the radio source may represent the detection of non-thermal emission produced by a magnetar central engine.  This would implicate magnetars as the energy sources powering SLSNe, as has been argued based on optical data (photometry and spectroscopy). Moreover, the radio source may be analogous to the persistent radio source associated with the repeating \repeater{}, thereby connecting these two classes of events \citep{Metzger2017}.  Indeed, we find that given the ejecta and engine parameters inferred from modeling of the optical data for PTF10hgi, the ejecta would be transparent to free-free absorption at the time of our observations.  In addition, scaling the model for the \repeater{} persistent source can reproduce the timescale and luminosity of the observed emission.  This model (and its details) can be further tested with multi-frequency radio observations, continued monitoring of the source brightness, and a search for scintillation-induced variability. 

We note that although our search for FRBs from the location of PTF10hgi yielded no detections, the expected probability of a detection is low in such a short duration observation (40 min); a more significant time investment with the GBT or Arecibo may yield detections or interesting limits.

\acknowledgments \textit{Acknowledgments.} We thank Joel Leja for helpful discussions about modeling of the host galaxy. The Berger Time-Domain Group at Harvard is supported in part by the NSF under grant AST-1714498 and by NASA under grant NNX15AE50G. SC and JMC acknowledge support from the NSF under grant AAG 1815242. RL acknowledges support from a Marie Sk\l{}odowska-Curie Individual Fellowship within the Horizon 2020 European Union (EU) Framework Programme for Research and Innovation (H2020-MSCA-IF-2017-794467). MN is supported by a Royal Astronomical Society Research Fellowship. The VLA observations presented here were obtained as part of program VLA/17B-171, PI: Berger. The VLA is operated by the National Radio Astronomy Observatory, a facility of the National Science Foundation operated under cooperative agreement by Associated Universities, Inc. This paper makes use of the following ALMA data: ADS/JAO.ALMA\#2017.1.00280.S. ALMA is a partnership of ESO (representing its member states), NSF (USA) and NINS (Japan), together with NRC (Canada), NSC and ASIAA (Taiwan), and KASI (Republic of Korea), in cooperation with the Republic of Chile. The Joint ALMA Observatory is operated by ESO, AUI/NRAO and NAOJ. The National Radio Astronomy Observatory is a facility of the National Science Foundation operated under cooperative agreement by Associated Universities, Inc. Support for program \#GO-15140 was provided by NASA through a grant from the Space Telescope Science Institute, which is operated by the Associations of Universities for Research in Astronomy, Incorporated, under NASA contract NAS5-26555. Support for this work was provided by NASA through the NASA Hubble Fellowship grant \#HST-HF2-51412.001-A awarded by the Space Telescope Science Institute, which is operated by the Association of Universities for Research in Astronomy, Inc., for NASA, under contract NAS5-26555.

\software{Boxfit \citep{vanEerten2012}, CASA \citep{McMullin2007}, CLOUDY \citep{Ferland2013}, Prospector \citep{prospector}, pwkit \citep{Williams2017}}

\bibliography{ref}
\bibliographystyle{apj}

\end{document}